\documentclass[twocolumn]{aastex631}
\makeatletter
\let\frontmatter@title@above=\relax
\makeatother
\usepackage{lipsum}

\usepackage{natbib}
\usepackage[english]{babel}
\usepackage[document]{ragged2e}
\usepackage{amsfonts}
\usepackage{amsmath}
\usepackage{amssymb}
\usepackage{bm}
\usepackage{braket}
\usepackage{enumitem}
\usepackage{float}
\usepackage{graphicx}
\usepackage{hyperref}
\usepackage{mathtools}
\usepackage{orcidlink}
\usepackage{soul}
\usepackage{url}
\usepackage{verbatim}
\usepackage{xfrac}
\usepackage{xspace}

\newcommand{\qbird}{{\sc qBIRD}\xspace}
\newcommand{\pycbc}{{\sc PyCBC}\xspace}
\newcommand{\bilby}{{\sc Bilby}\xspace}
\newlist{myitemize}{itemize}{1}
\setlist[myitemize,1]{    label=\textbf{\textbullet},left=0em,labelsep=1em,itemsep=1em,align=left,}

\begin{document}

\title{Quantum Bayesian Inference with Renormalization for Gravitational Waves}

\author{{Gabriel Escrig}\orcidlink{0000-0003-2881-085X}\,\hyperlink{email_addresses_start}{$^{\ast}$}}
\affiliation{Departamento de Física Teórica, Universidad Complutense de Madrid, 28040 Madrid, Spain}

\author{{Roberto Campos}\orcidlink{0000-0002-2527-4177}\,\hyperlink{email_addresses_start}{$^{\dagger}$}}
\affiliation{Departamento de Física Teórica, Universidad Complutense de Madrid, 28040 Madrid, Spain}
\affiliation{Quasar Science Resources, S.L., 28232 Madrid, Spain}

\author{Hong Qi\orcidlink{0000-0001-6339-1537}\,\hyperlink{email_addresses_start}{$^{\ddagger}$}}
\affiliation{School of Mathematical Sciences, Queen Mary University of London, London E1 4NS, UK}

\author{{M. A. Martin-Delgado}\orcidlink{0000-0003-2746-5062}\hyperlink{email_addresses_start}{$^{\S}$}}
\affiliation{Departamento de Física Teórica, Universidad Complutense de Madrid, 28040 Madrid, Spain}
\affiliation{CCS-Center for Computational Simulation, Universidad Politécnica de Madrid, Spain}

\justifying

\begin{abstract}
Advancements in gravitational-wave interferometers, particularly the next generation, are poised to enable the detections of orders of magnitude more gravitational waves from compact binary coalescences. While the surge in detections will profoundly advance gravitational-wave astronomy and multimessenger astrophysics, it also poses significant computational challenges in parameter estimation. 
In this work, we introduce a hybrid quantum algorithm \qbird, which performs quantum Bayesian Inference with Renormalization and Downsampling to infer gravitational wave parameters. We validate  the algorithm using both simulated and observed gravitational waves from binary black hole mergers on quantum simulators, demonstrating that its accuracy is comparable to classical Markov Chain Monte Carlo methods. Currently, our analyses focus on a subset of parameters, including chirp mass and mass ratio, due to the limitations from classical hardware in simulating quantum algorithms. However, \qbird can accommodate a broader parameter space when the constraints are eliminated with a small-scale quantum computer of sufficient logical qubits.

\end{abstract}

\justifying

\section{Introduction}

The Advanced LIGO and Advanced Virgo observatories \citep{LIGOScientific:2014pky, VIRGO:2014yos} have detected about 100 gravitational waves (GWs) from compact binary coalescences \citep{LIGOScientific:2018mvr, LIGOScientific:2020ibl, KAGRA:2021vkt} since their first observation of a binary black hole (BBH) merger in 2015 \citep{abbott2016}. A network of third-generation (3G) gravitational-wave observatories, such as Cosmic Explorer (CE) \citep{PhysRevD.91.082001,PhysRevD.96.084004,PhysRevD.96.084039,hall2020,Evans:2021gyd}, Einstein Telescope (ET) \citep{punturol2010}, and Neutron Star Extreme Matter Observatory (NEMO) \citep{ackley2020} will significantly advance our capacity in detecting GWs, including those from compact binary coalescences, core-collapse supernovae, and rotating compact objects \citep{Kalogera:2021bya}. Consequently, GW source parameter estimation will face unprecedented computational challenges \citep{Couvares:2021ajn}. Moreover, in an era of thousands of detections per day \citep{Gupta:2023lga}, the majority of the signals are overlapped. With the sensitivity improvements in low frequency band, signals can be tracked at lower frequencies and over much longer durations, extending from currently seconds to hours. These challenges cannot be easily addressed by traditional parameter estimation tools. 

Quantum computing emerges as a promising solution for GW data analysis challenges, despite its rarity in gravitational-wave astronomy research. In particular, recent works have shown proof-of-principle applications of quantum algorithms in GW detection \citep{PhysRevResearch.4.023006,Miyamoto:2022pmh,Veske:2022tmh,Hayes:2023buy}. Our exploration focuses on GW source parameter estimation with a prioritization of accuracy and precision, recognizing that while computation speed remains a challenge for classical methods, current quantum technology has not yet matured enough to demonstrate an execution speed advantage in analysis. Quantum techniques are particularly useful for search and sampling problems \citep{hangleiter2023}. In a previous study \citep{escrig2023}, we proved a quantum polynomial scaling advantage over classical algorithms in ranking GW likelihoods we now retain this scaling advantage while enhancing the efficiency of the algorithm. In this work, we develop a comprehensive computational framework that implements a quantum version of the classical Markov Chain Monte Carlo (MCMC) technique \citep{vandersluys2008}, specifically, its archetype, the Metropolis-Hastings (MH; also referred to as Metropolis for convenience) algorithm \citep{christensen2004} to compute posterior probability density functions (PDFs) of GW source parameters, achieving accuracy comparable to classical methods \citep{usman2016, Biwer:2018osg, ashton2019}. 

We present \qbird, a hybrid quantum algorithm for gravitational wave (GW) source characterization that incorporates Bayesian inference enhanced by renormalization and downsampling. We demonstrate its capabilities by inferring both simulated and observed gravitational wave signals from merging BBHs. The analysis results show that \qbird achieves accuracy and precision comparable to classical methods. Additionally, it provides high sampling precision and efficient state-space representation in superposition with minimal qubits, requiring fewer iterations for the algorithm to converge.  Furthermore, \qbird exhibits broader scalability and the potential to outperform classical techniques as quantum hardware continues to advance.

\section{Algorithms}
\subsection{GW likelihood}
For a detected gravitational wave, Bayesian inference is applied to characterize source properties.   
Given data $d$ and model $M$, characterizing the parameter space $\boldsymbol{\theta}$ that models a gravitational wave signal $h(\boldsymbol{\theta})$ is estimating the posterior probability density functions (PDFs) $p(\boldsymbol{\theta} | d, M)$.  Bayes' theorem yields these posteriors as

\begin{equation}\label{bayes}
\small
p(\boldsymbol{\theta} | d, M) = \frac{\pi(\boldsymbol{\theta} | M) {\cal L}(d | \boldsymbol{\theta}, M)}{Z_{M}},
\end{equation}
where $\pi(\boldsymbol{\theta} | M)$ is the prior probability that models the belief in $\boldsymbol{\theta}$ under $M$, ${\cal L}(d | \boldsymbol{\theta}, M)$ is the gravitational wave likelihood representing the probability of observed data $d$ given the parameters $\boldsymbol{\theta}$ and model $M$, and  $Z_{M} = \int {\cal L}(d | \boldsymbol{\theta}, M) \pi(\boldsymbol{\theta} | M) d \boldsymbol{\theta}$ is the normalization constant for the marginalized posterior likelihood, or evidence. The inference process involves computing and ranking the likelihoods between gravitational wave signals $h(\boldsymbol{\theta})$ predicted by theory and the noisy observed data $d$. 
Since the noise (the difference between observed data and GW signals) is assumed to be stationary and Gaussian, and it is characterized by the power spectral density (PSD), $S_n$, the GW likelihood follows a Gaussian about the square root of the PSD. In the frequency domain, it is
\begin{equation}\label{likelihood}
\small
{\cal L}(d|\text{\boldmath$\theta$}, M) \propto \text{exp}\left( -\frac{1}{2}\sum_{i=1}^{N}\frac{|{d}(f_{i}) - {h}(f_{i};\text{\boldmath$\theta$})|^{2}}{S_{n}(f_{i}) }\right),
\end{equation}
where $N$ is the total number of frequency nodes $ f_{i}$.

The GW likelihood in Equation \eqref{bayes} has two important properties for our purposes of constructing a hybrid quantum algorithm for parameter estimation based on renormalization methods; see Step 1 in Section \ref{subsec:qbird} later: i) given a stationary PSD, the GW likelihood depends solely on the disparity between the model and observed data, representing pure noise as long as the model aligns with the observed data, and is the product of the individual frequency bins likelihoods, and ii) the priors for source property are independent:
\begin{equation}\label{prior}
\small
\pi( \text{\boldmath$\theta$} |M) = \prod_{p=1}^{P} \pi(\theta_{p} |M),
\end{equation}
with $ P $ the total number of parameters to infer. These factorization properties are the foundation for truncation (renormalization) in the quantum space of states that represent all parameters, leading to the formulation of our algorithm, which is presented and employed in the following sections.

\begin{figure}[t]
    \includegraphics[width=0.45\textwidth]{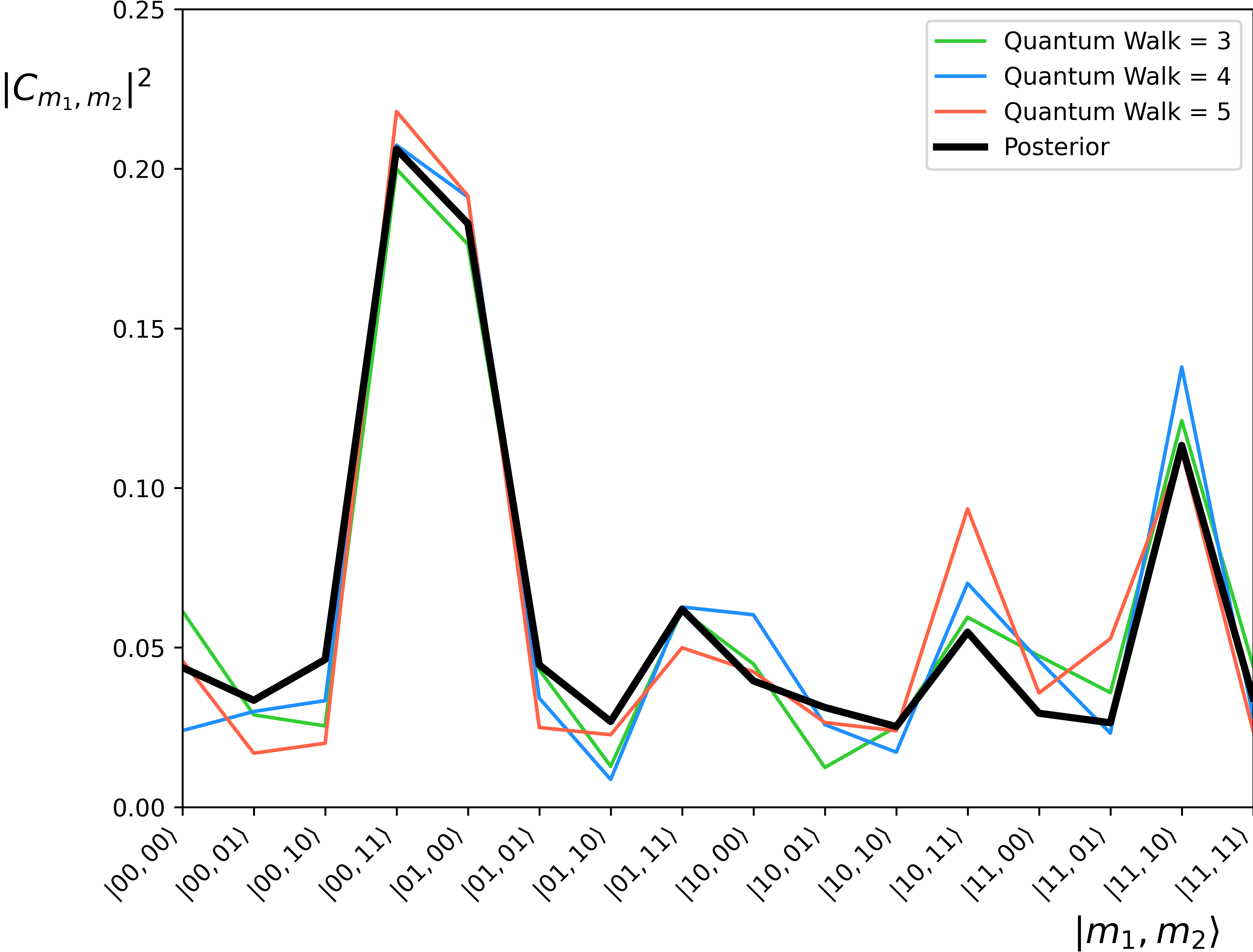}
    \caption{\label{fig:sampling}\justifying Discrete posterior probabilities for 16 combinations of the two component masses for GW150914 using $Q = 2$ discretization qubits. The colored lines show the state register probabilities $\ket{\bm{\theta}}_S = \ket{m_1, m_2}_S$ for different numbers of applications of the quantum walk operator in Equation (\ref{eqn:EvOp}). The black line represents the posterior probabilities in Equation (\ref{bayes}) for the 16 component mass combinations.}
\end{figure}

\subsection{Quantum Metropolis algorithm for GW likelihood ranking}
Parameter estimation from GW data uses stochastic sampling techniques to draw samples from the posterior distributions as described in Equation \eqref{bayes}. In GW community, this statistical analysis predominantly relies on MCMC methods \citep{christensen2022}, which demand computationally intensive numerical methods and high-performance computers. This opens the door for applying quantum algorithms developed in this work. Several approaches have been proposed to extend classical Metropolis algorithms into the quantum domain \citep{temme2011}, showing an anticipated quantum advantage over their classical counterparts \citep{yung2012, lemieux2020}. 
In this work, we introduce a hybrid Metropolis heuristic algorithm based on quantum walks called \qbird, quantum Bayesian Inference with Renormalization and Downsampling. This approach enables the inference of source properties for gravitational waves in both observed data in the LIGO O1 observing run and data comprising injected waveforms into a Gaussian noise, which we call injections. It not only showcases the quantum computational advantages but also demonstrates comparable inference accuracy to classical methods \citep{ashton2019, usman2016}. 

A quantum walk can be viewed as an agent which explores the parameter space in superposition \citep{aharonov2001}, which is endowed with a quantum Hilbert space \citep{szegedy2004} of states specified as follows. Let $ \Theta $ be the configuration space of the parameters $\text{\boldmath$\theta$}$ we want to infer with the experimental data. The dimensionality of this space depends on the total number of parameters, about 15 to 20 for a typical compact binary merger event. $ \Theta $ must be discretized with a certain grid or lattice that also depends on the precision used to represent each parameter $ \theta_{p} $. Our choice is a hypercube lattice with periodic boundary conditions that allow quantum walks between nearest-neighbor vertices.

To specify a quantum walk in this space state of parameters, we use 3 quantum registers, similar to other quantum walk proposals \citep{lemieux2020, miyamoto2023}. First, a register of states $ \ket{\text{\boldmath$\theta$}}_{S} $ stores the information of the parameter values. A second register $\ket{p}_{D}$ encodes the hopping directions of the walker in binary notation corresponding to the oriented edges of the lattice. A third register $ \ket{\Delta \text{\boldmath$\theta$}}_{E} $ stores the information of, given parameter $ \theta_{p} $, moves to a neighbor site by shiftting the parameter an amount $ \Delta \theta_{p} $ or an amount $ - \Delta \theta_{p} $. Additionally, a coin state $ \ket{\varphi }_{C} $ accounts for the random evolution of the walker. Finally, an auxiliary register $\ket{A(\bm{\theta}, \bm{\theta}+ \Delta \bm{\theta})}_{A}$ stores the acceptance probabilities of each transition. These are given by the MH acceptance rule: 
\begin{equation}\label{acceptance}
        \small
        A(\bm{\theta}, \bm{\theta}+ \Delta \bm{\theta}) = \min [ 1,  \frac{\pi(\bm{\theta}+ \Delta \bm{\theta})}{\pi(\bm{\theta})} ( \frac{\mathcal{L}(d| \bm{\theta}+ \Delta \bm{\theta})}{\mathcal{L}(d| \bm{\theta})})^\beta ],
    \end{equation}
where $\beta$ represents an annealing schedule. 

The quantum walk employs a total of $PQ + \lceil \log_2 P \rceil + a + 2 $ qubits: $PQ$ represents the number of qubits needed for the register $\ket{\text{\boldmath$\theta$}}_{S}$ that contains all the points of the lattice $\Theta$, where $P$ is the number of inferred parameters and $Q$ is the number of discretization qubits, with $2^Q$ states represented for each parameter; $\lceil \log_2 P \rceil$ qubits to represent the register $\ket{p}_{D}$ in binary encoding; $a$ qubits to represent the auxiliary register for acceptance probability. Finally, 2 qubits are needed, one for the register $ \ket{\Delta \text{\boldmath$\theta$}}_{E}$, and another for the coin register $ \ket{\varphi }_{C}$ to encode the accept/reject probability of all states.

\begin{figure*}[t]
  \centering
  \includegraphics[width=0.98\textwidth]{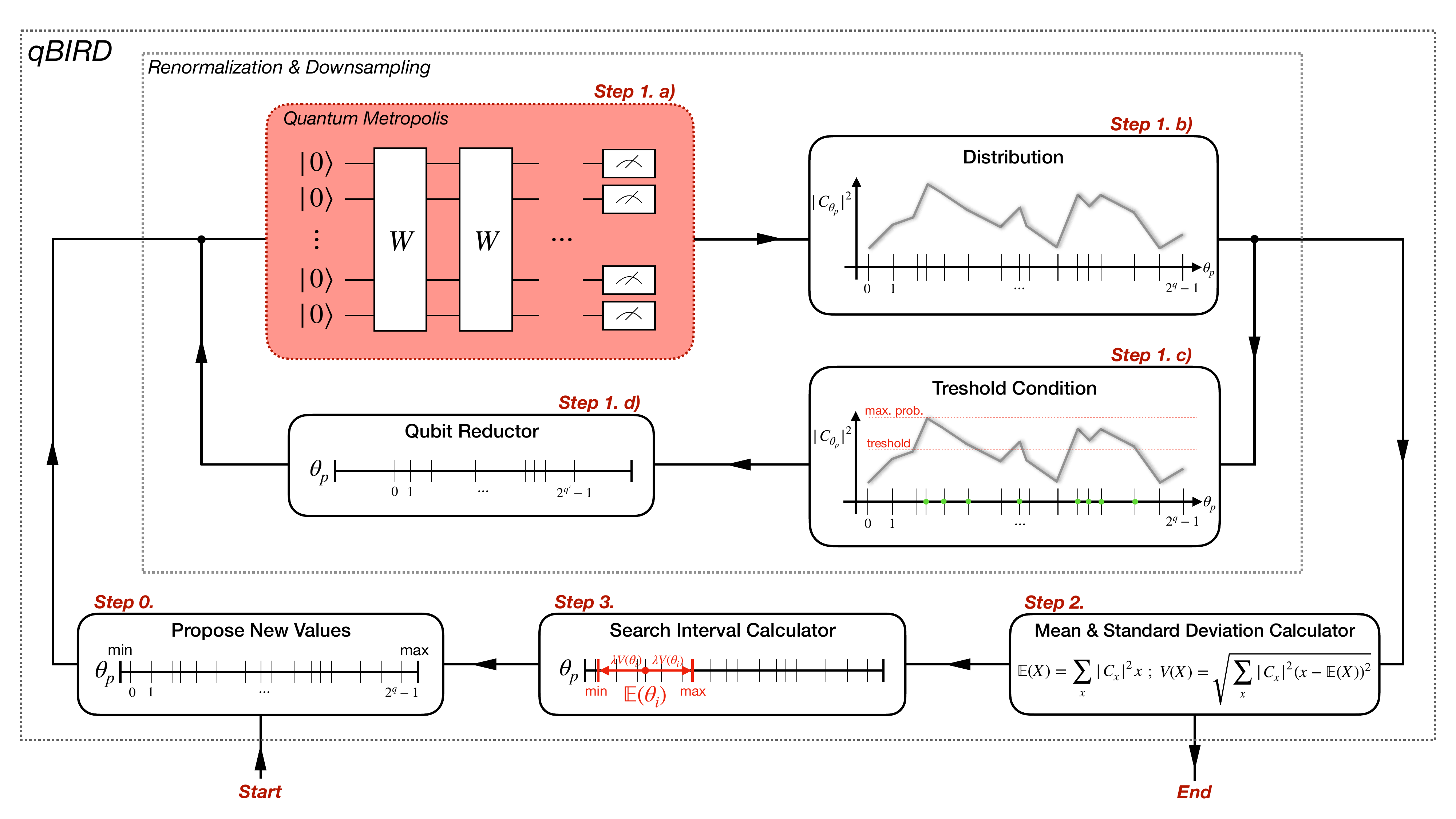}
    \caption{\label{fig:qBird_Schedule} Flowchart of \qbird algorithm. See main text for module and step explanations.}
\end{figure*}

Now, the evolution operator $W$ of the quantum walk is constructed over the previous registers as follows (see \ref{sm_quantumwalk} for its detailed construction):
\begin{equation} \label{eqn:EvOp}
    W = R V^{\dagger} B^{\dagger} S F B V.
\end{equation}

This enables the construction of the one-step circuit of the quantum walk, with $W$ as a building block of a quantum MH algorithm. By applying several $W$'s consecutively, a quantum walk traverses the parameter space $ \Theta $ according to certain transition probabilities, just as the classical MH does. The operator $W$  samples the posterior distribution in Equation \eqref{bayes}, storing it as state probabilities, as shown in Figure \ref{fig:sampling}. The crucial aspect underpinning the superiority of the quantum algorithm over its classical counterpart lies in that each measurement simulates the probability of accepting each state, capitalizing on the efficiency of superposition. Prior research demonstrated that the quantum MH algorithm achieved a polynomial scaling advantage over its classical counterpart for GW likelihood ranking \citep{escrig2023}.

\subsection{\qbird algorithm}
\label{subsec:qbird}

Using the above quantum walk as a core, we have developed a hybrid algorithm, \qbird, capable of inferring posterior probability-density functions of the source parameters of gravitational waves from BBH mergers. In addition to compact binary mergers, this algorithm is applicable to modeled waveforms from any types of gravitational-wave sources, including those from core-collapse supernovae. \qbird consists of three modules: the quantum Metropolis module, the renormalization and downsampling module, and a classical postprocessing module. The description of the algorithm is as follows, and is illustrated schematically in Figure \ref{fig:qBird_Schedule}.

\noindent \textbf{Step 0.} \textbf{Parameter initialization:} The algorithm is initialized by proposing $2^Q$ values for each parameter, drawn from a uniform distribution specified by the lower and upper bounds of the prior function given by Equation \eqref{prior}. All these values are stored in the state register $\ket{\text{\boldmath$\theta$}}_{S}$ producing an initial state $ \ket{\phi^{(0)}} $. 

\noindent \textbf{Step 1.} \textbf{Renormalization and downsampling: } This module executes the quantum Metropolis algorithm in Equation \eqref{eqn:EvOp}, which is adapted from \citep{Campos2023} and endowed with a renormalization method that defines the \qbird algorithm. In this step, the quantum walk is applied several times as we decrease the parameter space $\Theta$ to locate the set of values for each parameter $\theta_p$ maximizing the likelihood.

\noindent   \textbf{a) Quantum Metropolis: }
Iteratively apply the walk operator in Equation \eqref{eqn:EvOp} for $L$ times on the initial state, which contains $|S| := s = PQ$ qubits:
        \begin{equation} \label{eqn:ev_state}
            \ket{\psi (L)} := W_L ... W_2 W_1 \ket{\phi^{(0)}}.
        \end{equation}
The integer $s$ is also used as the index of the renormalization procedure.

\noindent   \textbf{b) Distribution:} Get the probability distribution of the state register $\ket{\text{\boldmath$\theta$}}_{S}$,
        \begin{equation}
            \small
            \ket{\bm{\theta}}_S := \sum_{x \in \Theta (s)} C_x \ket{x}_S,
        \end{equation}
from $\ket{\text{\boldmath$\theta$}}_{S}$ measurements to obtain the pairs $\{ |C_x|^2, x \}_s$. $\Theta (s)$ denotes the state space of qubits at the $s$-th step of the renormalization to be described in Step 1.d).

\noindent  \textbf{c) Threshold condition:} If $s = P$,  jump to Step 2; otherwise calculate the number of elements $ |S_{h}(s)| $ with
        \begin{equation}
        \small
            S_h(s) := \left\{ y \in \Theta(s) : |C_y|^2 \geq \alpha \underset{x \in \Theta(s)}{\max} |C_x|^2 \right\},
        \end{equation}
        where $\alpha \in [0, 1]$ represents a threshold, and $ S_{h}(s) $ is a sieve or filter function to obtain it.
                    
 \noindent \textbf{d) Qubit reductor:} Reduce the number of qubits in the state register $\ket{\text{\boldmath$\theta$}}_{S}$ by defining
        \begin{equation}
        \small
            s^\prime := \max \left[ P , \min \left( \lceil \log_2 |S_h (s)| \rceil , s - P \right) \right],
        \end{equation}
        and go to Step 1.a) with $s^\prime$ qubits and the $2^{s^\prime}$ highest probability values. This condition enables the elimination of at least one qubit for each parameter, ending up with a minimum of one qubit per parameter.

The second module arises from the challenge of using the quantum Metropolis algorithm to search for the state with the maximum probability and is inspired by the renormalization techniques of quantum lattice models  \citep{wilson1975}. Due to the enormous size of the state space, the normalization factor of quantum states results in very small probability differences between the most and least probable states. Although the probability disparity between states may span a couple of orders of magnitude, obtaining significance would require an impractical number of measurements.  During the discretization process, the evidence in Equation \eqref{bayes} is proportional to the size of the lattice, $Z_M \propto | \Theta |$. Then, as we increase the size of the parameter space $\Theta$, the closer to zero the probabilities will be. It is important to note that this problem is specific to Bayes' theorem and has not been introduced by using quantum computing. 
However, if we gradually remove the states that are significantly less probable by reducing the size and qubits of the problem, these differences become progressively more noticeable. With this technique, we are able to find the state with the maximum likelihood over all the proposed values.

The effectiveness of the quantum renormalization method in computing the maximum likelihood, shown in Equation (\ref{bayes}), stems from the well-suited truncations in Hilbert space of states for uncorrelated noise (Gaussian) describing the likelihood in Equation \eqref{likelihood} and the black hole parameters in Equation \eqref{prior}, as exemplified by properties i) and ii) after Equation (\ref{likelihood}).

This technique also makes the algorithm more robust, being able to discern the correct states in more detail even as the state of the system grows, thus decreasing the error over iterations. Moreover, it reduces the circuit depth, shortening it only to the depth of the number of W operators applied.

\begin{figure}[t]
\centering
        \includegraphics[width=0.45\textwidth]{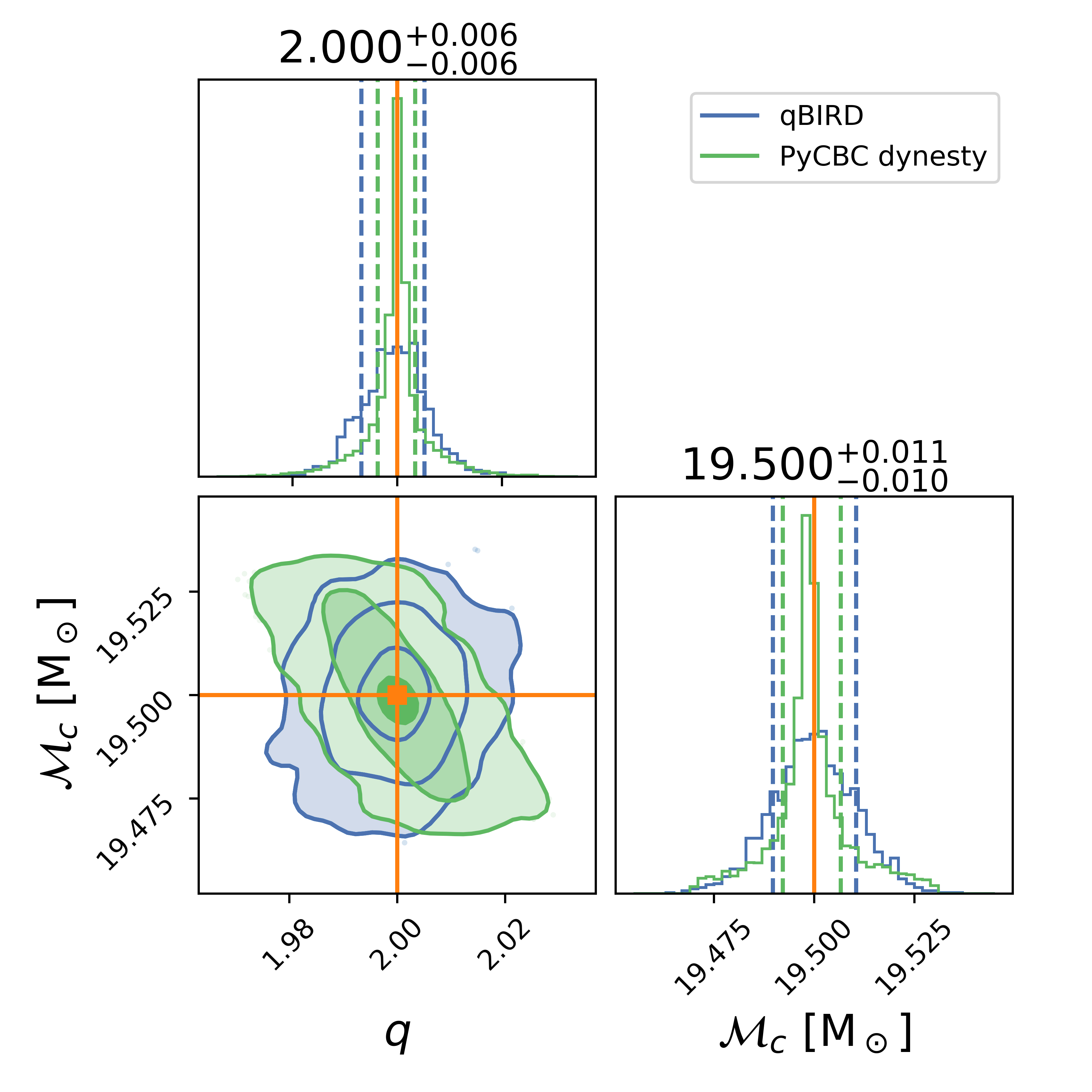}
\caption{Comparison of posterior distributions of  chirp mass $\mathcal{M}_c$ and mass ratio $q$ obtained using \qbird (blue) and \pycbc \textit{dynesty} sampler (green) for a simulated BBH gravitational-wave signal injected into Gaussian-noise using \pycbc. The injected values (organge) are $\mathcal{M}_c = 19.5\ {\text{M}_{\odot}}$ and $q = 2$.}
    \label{fig:injections_2param}
\end{figure}

\noindent  \textbf{Step 2.} \textbf{Mean and standard deviation calculator:} The third module consists of a classical processing that takes the results obtained in the first two modules to generate posteriors for each parameter and converge the algorithm. Hence, given the pairs, $\{ |C_x|^2, x \}_{s = P}$ compute
        {\small
        \begin{align}
            \mathbb{E} (\theta_p)  &:= \sum_{x \in \theta_p(s = P)} |C_x|^2 x, \\
            V(\theta_p) &:= \sqrt{\sum_{x \in \theta_p(s = P)} |C_x|^2 (x - \mathbb{E} (\theta_p))^2},
        \end{align}}
\noindent{which represent the mean and weighted standard deviation for each parameter $p = 1, \dotsb , P$, respectively. It is important to save the $\mathbb{E} (\theta_p)$ values in each iteration in order to build the PDFs at the end of the algorithm.}

 \noindent  \textbf{Step 3.} \textbf{Search interval calculator:} To gradually narrow down the search area, a new interval for each parameter is proposed from $\mathbb{E} (\theta_p)$ and $V(\theta_p)$  previously obtained, with lower and upper values given by:
    \begin{equation}
            \theta_{p, (\text{min},\text{max})}  = \mathbb{E} (\theta_p) \mp  \lambda V(\theta_p),    
     \end{equation}
where $\lambda$ is a parameter to be set for controlling the convergence of the algorithm. Note that the proposed new minimum (maximum) cannot be lower (greater) than the one set by prior interval (\ref{prior}). Then return to Step 0 with the new interval $[\mathbb{E} (\theta_p) - \lambda V(\theta_p), \mathbb{E} (\theta_p) + \lambda V(\theta_p)]$.

\noindent    \textbf{End.} 
After a given number of iterations of Steps $0 - 3$, posterior distributions of each of the parameters $\theta_p$ are constructed from the $\mathbb{E} (\theta_p)$ values obtained in each iteration.

\begin{figure}[t]
        \centering       \includegraphics[width=0.498\textwidth]{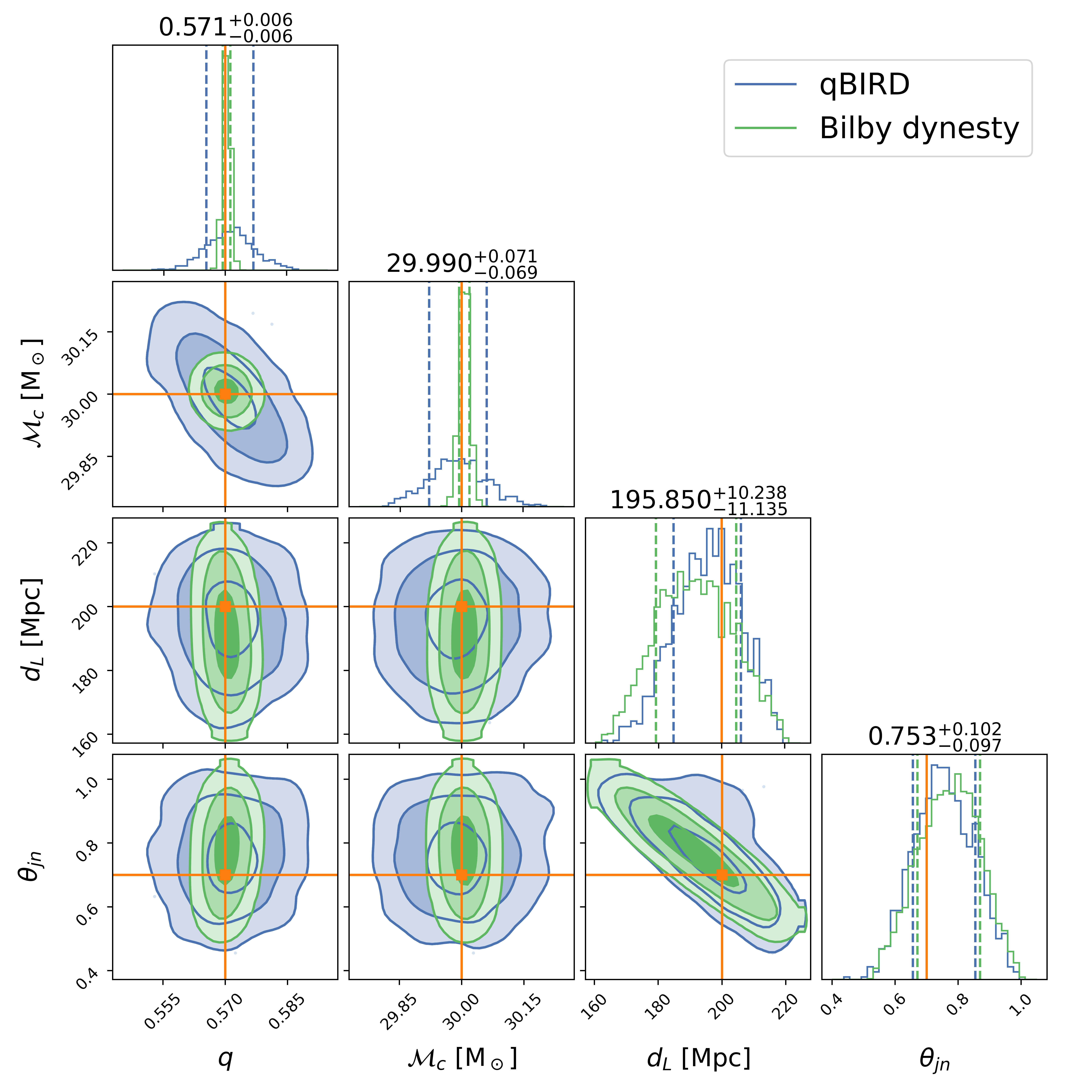}
    \caption{\justifying Comparison of posterior distributions obtained using \qbird (blue) and \bilby \textit{dynesty} sampler (green) for a simulated BBH gravitational-wave signal characterized with 4 parameters injected into zero noise using \bilby. The injected values (orange) are $\mathcal{M}_c=30\ {\text{M}_{\odot}}$, $q=0.57$, $d_L=200 \text{ Mpc}$, and $\theta_{jn}=0.7\ \text{radian}$.}
    \label{fig:injections_4param}
\end{figure}

\section{Applications and Results}
\label{sec:results}
As a demonstration and validation, we applied \qbird to analyze two simulated gravitational waves from BBH mergers, and compared the recovered posteriors with the known injected parameter values and those obtained using classical parameter estimation pipelines \pycbc and \bilby. Throughout this work, the IMRPhenomPv2 model \citep{Khan:2018fmp} is used for both signal injections and parameter recoveries. In addition, we applied the \qbird algorithm to the LIGO O1 observed data; refer to in Appendix \ref{extra_res}.

The first simulated gravitational wave has 2 unknown parameters: chirp mass $\mathcal{M}_c$ and mass ratio $q$. It was injected into a Gaussian noise \citep{gaussiannoise44} using \pycbc \citep{alex_nitz_2024_10473621}, with injected values $\mathcal{M}_c = 19.5\ {\text{M}_{\odot}}$ and $q = 2$. In the parameter estimation process, the priors are uniformly distributed $\mathcal{M}_c \in [19.4, 19.6] \ {\text{M}_{\odot}}$ and $q \in [1.9, 2.1]$ for both \qbird and \pycbc. For \qbird, $Q = 6$ discretization qubits were used for each parameter, and 2100 iterations were performed,  yielding one posterior sample point per iteration, with the first 100 burn-in points discarded. Convergence was observed after the first tens of iterations due to the benefit of quantum superposition, but 2000 more were used to smooth the posteriors. Each parameter was sampled by $2^6$ discrete points, yielding $2^{12}$ possible combinations for the two parameters. A total of 18 qubits were used to execute the circuit, which performed 4 steps of $W$ in each iteration with a constant annealing schedule of $\beta = 0.5$. The \pycbc \textit{dynesty} sampler was performed with the following settings: $\text{nlive} = 1000$, $\text{dlogz} = 0.1$, and $\text{max\_iter}=5000$, which resulted in 11813 posterior sample points, of which the first 4010 were discarded to ensure only converged points were kept. Figure \ref{fig:injections_2param} shows the comparison of the inference results using \qbird and the \pycbc \textit{dynesty} sampler, respectively, in estimating the chirp mass $\mathcal{M}_c$ and mass ratio $q$ of the first simulated gravitational wave. \qbird recovered the injected values $\mathcal{M}_c = 19.5\ {\text{M}_{\odot}}$ and $q = 2$ with accuracy and precision comparable to \pycbc. The inferred peak value of each parameter falls within the $90\%$ confidence interval of its posterior, and the measurement uncertainties are tiny fractions of the injected values. The slight derivations of the peak values from the injected values and the small uncertainties are caused by the Gaussian noise. 

The second simulated BBH merger has 4 parameters: chirp mass $\mathcal{M}_c$, mass ratio $q$, luminosity distance $d_L$, and inclination angle $\theta_{jn}$, constructed using \bilby \citep{ashton2019} into zero noise, with injected values $\mathcal{M}_c=30\ {\text{M}_{\odot}}$, $q=0.57$, $d_L=200 \text{ Mpc}$, and $\theta_{jn}=0.7\ \text{radian}$. 
The recoveries were performed using \qbird and \bilby \textit{dynesty}, with the priors being $\mathcal{M}_c \in [27, 35] \ {\text{M}_{\odot}}$, $q \in [0.25, 1]$,  $d_L \in [150, 220] \text{ Mpc}$, and $\theta_{jn} \in [0, 1]\ \text{rad}$ for both. 
Due to the limitations of quantum simulators on classical computers, the number of discretization qubits for each parameter was reduced to $Q = 3$ and the parameter estimation was conducted with 1200 iterations using \qbird, with the first 100 burn-in iterations discarded. Each parameter was sampled by $2^3$ points, with $2^{12}$ possible combinations for the four parameters. A total of 19 qubits were used to execute the circuits, which performed 4 steps of $W$ in each iteration with a constant annealing schedule of $\beta = 0.05$. The settings for \bilby \textit{dynesty} are:  $\text{nlive} = 1000$ and $\text{dlogz} = 0.1$, which resulted in 3464 sample points after convergence. Figure \ref{fig:injections_4param} shows the comparison of posterior distributions between \qbird and the \bilby \textit{dynesty} sampler for the 4-parameter case. Although this case is significantly complicated due to more parameters being estimated and fewer discretization qubits being used for each parameter, \qbird was still able to reproduce the injected values accurately. The measurement uncertainties are larger than \bilby because fewer qubits were used for each parameter, which is a limit set by classical hardware in simulating quantum circuits. Simulating more qubits for each parameter requires deeper circuits which will take much longer time to compile. However, the 2-parameter has demonstrated \qbird's capabilities in precision. This 4-parameter estimation instance further demonstrates the scalability potential of \qbird in estimating more parameters when a robust quantum environment becomes feasible.  

We now move to a discussion of algorithm benchmarks. qBIRD is a hybrid quantum algorithm, with quantum approaches applied in the upper loop of the workflow (the likelihood ranking) and part of the sampling process, as shown in Figure \ref{fig:qBird_Schedule}. Given the early stages of quantum hardware and the limitations of simulating quantum circuits on classical hardware, runtime is not yet comparable to that of classical analysis methods. To evaluate performance, we use the Total Time to Solution (TTS) metric, previously applied in our earlier works \citep{Campos2023, escrig2023}. This metric represents the expected average steps required for the algorithm to find a solution, assuming that the procedure is repeated multiple times. Notably, the quantum approach demonstrates a polynomial advantage over classical methods when dealing with large-scale likelihoods and complex likelihood distributions.

\section{Conclusions}
We have introduced \qbird, our hybrid quantum algorithm for GW parameter estimation, and showcased its accuracy and precision by applying it to both 2-parameter and 4-parameter simulated gravitational waves from binary black hole mergers. This algorithm builds upon the quantum Metropolis solver framework \citep{Campos2023}, which has demonstrated a scaling advantage for GW likelihood ranking \citep{escrig2023}. By introducing renormalization and downsampling techniques with quantum walks, \qbird realizes a quantum version of the classical MCMC sampler. The posterior distributions inferred using \qbird for the 2- and 4-parameter example BBHs in Section \ref{sec:results} are accurate, precise, and consistent with the results obtained with classical computing methods. 

The limited number of parameters is a compromise between the largest possible parameter size and the capabilities of quantum simulators running on classical hardware to obtain the inference results. Quantum simulators on classical computers are inherently limited in the depth of circuits and the number of qubits they can execute, and the same limitations apply to existing quantum hardware. Therefore, in showcasing our algorithm, we only used up to 4 parameters. However, since \qbird is scalable, it will benefit from the upward trend of advancements in quantum technology in the near term.

Although quantum computing is still in its early stages, this work represents a significant step forward as the first quantum hybrid sampler for gravitational wave parameter estimation. It has demonstrated clear advantages in gravitational-wave likelihood ranking and sampling and inspired confidence in its potential for a full quantum implementation of the parameter estimation workflow.

\section{Acknowledgements}
G.E., R.C., and M.A.M.-D. acknowledge the support from grants MINECO/FEDER Projects, PID2021-122547NB-I00 FIS2021, MADQuantumCM project funded by Comunidad de Madrid, the Recovery, Transformation, and Resilience Plan, NextGenerationEU, funded by the European Union, and the Ministry of Economic Affairs Quantum ENIA project funded by Madrid ELLIS Unit CAM. M.A.M.-D. has also been partially supported by the U.S. Army Research Office through Grant No.W911NF-14-1-0103. This work used the computing facilities of the GICC group at UCM. This work has been financially supported by the Ministry for Digital Transformation and of Civil Service of the Spanish Government through the QUANTUM ENIA project call – Quantum Spain project, and by the European Union through the Recovery, Transformation and Resilience Plan – NextGenerationEU within the framework of the Digital Spain 2026 Agenda.$\ \ $  H.Q. thanks Frank Linde for inspiring her quantum computing exploration for gravitational wave data analysis in 2021 summer. H.Q. extends gratitude to Gabriela Gonzalez and her team at Louisiana State University and LIGO Livingston Observatory for their hospitality during this work in 2023.  H.Q. was supported in part by the 2022-2023 additional STFC IAA grant to Queen Mary University of London. H.Q. is grateful for computational resources provided by LIGO Laboratory and the Leonard E Parker Center for Gravitation, Cosmology and Astrophysics at the University of Wisconsin-Milwaukee and supported by National Science Foundation Grants PHY-0757058, PHY-0823459, PHY-1626190, and PHY-1700765. This material is based upon work supported by NSF's LIGO Laboratory which is a major facility fully funded by the National Science Foundation. 

G.E. and R.C. equally contributed to this work.

\hypertarget{email_addresses_start}{}
$^{\ast}$ \href{mailto:gescrig@ucm.es}{gescrig@ucm.es}

$^{\dagger}$ \href{mailto:robecamp@ucm.es}{robecamp@ucm.es}

$^{\ddagger}$ \href{mailto:hong.qi@ligo.org}{hong.qi@ligo.org}

$^{\S}$ \href{mailto:mardel@ucm.es}{mardel@ucm.es}

\newpage
\appendix

\section{Quantum Walk Operator Construction} \label{sm_quantumwalk}
The quantum walk operator $W = R V^{\dagger} B^{\dagger} S F B V$ in Equation \eqref{eqn:EvOp} is composed of the following elementary operations.
It starts by making a superposition over all possible movements with the $ V $ operator,

\begin{equation}
    V \ket{0}_D \ket{0}_E = \frac{1}{\sqrt{2p}} \sum_{i = 0}^{p-1} \ket{i}_D  \sum_{j \in \{ 0 , 1\}}  \ket{j}_E = \frac{1}{\sqrt{2p}} [\ket{0}_D + \ket{1}_D + \cdots + \ket{p-1}_D] \otimes [\ket{0}_E + \ket{1}_E].
\end{equation}

It is implemented by applying Hadamard gates to all qubits. Once all possible moves are in superposition, the acceptance probabilities in Equation \eqref{acceptance} are encoded into the coin register with the $ B $ operator:
\begin{equation}
    B \ket{\bm{\theta}}_{S} \ket{i}_D  \ket{\Delta \text{\boldmath$\theta$}_i}_{E}  \ket{A(\bm{\theta}, \bm{\theta}+ \Delta \bm{\theta}_i)}_{A} \ket{\varphi}_C
     = \ket{\bm{\theta}}_{S} \ket{i}_D \ket{\Delta \text{\boldmath$\theta$}_i}_{E} \ket{A(\bm{\theta}, \bm{\theta}+ \Delta \bm{\theta}_i)}_{A}  {\cal U}(\vartheta) \ket{\varphi}_C ,
\end{equation}
where  $\Delta \bm{\theta}_i = (0, 0, \ldots, \stackrel{\text{parameter } i}{\Delta \theta}, \ldots, 0)$, each element of the vector being a specific parameter.  Implementing this operator consists of a rotation
$ {\cal U}(\vartheta) $ of angle $\vartheta = \arcsin \left( \sqrt{A(\bm{\theta}, \bm{\theta}+ \Delta \bm{\theta}_i)}\right) $ controlled by the $\ket{A(\bm{\theta}, \bm{\theta}+ \Delta \bm{\theta}_i)}_{A}$ register.
At this point, the transition in the state register $\ket{\bm{\theta}}_{S}$ is performed by the $ F $ operator:
\begin{equation}
    F \ket{\bm{\theta}}_{S} \ket{i}_D \ket{\Delta \text{\boldmath$\theta$}_i}_{E} \ket{\varphi}_C 
     = \begin{cases}
         \ket{\bm{\theta}}_{S} \ket{i}_D \ket{\Delta \text{\boldmath$\theta$}_i}_{E} \ket{0}_C & \text{if } 
         \ket{\varphi}_C = \ket{0}_C, \\
         \ket{\bm{\theta}+ \Delta \bm{\theta}_i)}_{S} \ket{i}_D \ket{\Delta \text{\boldmath$\theta$}_i}_{E} \ket{1}_C & \text{if } \ket{\varphi}_C = \ket{1}_C.
       \end{cases}
\end{equation}
and can be constructed from an adder gate conditioned by the coin register $\ket{\varphi}_C$.
Then, the operator $S$ flips the sign of the value in the register $\ket{\Delta \text{\boldmath$\theta$}_i}_{E}$ conditioned by the coin register $\ket{\varphi}_C$:
\begin{equation}
    S \ket{\bm{\theta}}_{S} \ket{i}_D \ket{\Delta \text{\boldmath$\theta$}_i}_{E} \ket{\varphi}_C 
     = \begin{cases}
         \ket{\bm{\theta}}_{S} \ket{i}_D \ket{\Delta \text{\boldmath$\theta$}_i}_{E} \ket{0}_C & \text{if } 
         \ket{\varphi}_C = \ket{0}_C, \\
         \ket{\bm{\theta}}_{S} \ket{i}_D \ket{- \Delta \text{\boldmath$\theta$}_i}_{E} \ket{1}_C & \text{if } \ket{\varphi}_C = \ket{1}_C.
       \end{cases}
\end{equation}
and can be constructed from a CNOT gate controlled by the coin register $\ket{\varphi}_C$.
Finally, the changes in the movement and coin registers are reversed and then the $\ket{0}_P \ket{0}_E \ket{0}_C$ state is subject to the following reflection with the $R$ operator defined as follows:
\begin{equation}
    R \ket{i}_D \ket{\Delta \text{\boldmath$\theta$}_i}_{E} \ket{\varphi}_C  
    = \begin{cases}
         - \ket{0}_D \ket{\text{\boldmath$0$}}_E \ket{0}_C & \text{if } (i,\Delta \text{\boldmath$\theta$}_i, \varphi) = (0, \text{\boldmath$0$}, 0), \\
         \ket{i}_D \ket{\Delta \text{\boldmath$\theta$}_i}_{E} \ket{\varphi}_C & \text{otherwise.}
       \end{cases}
\end{equation}

\section{Additional parameter estimation with \qbird} \label{extra_res}
\begin{figure*}[t]
  \centering
  \includegraphics[width=0.5\textwidth]{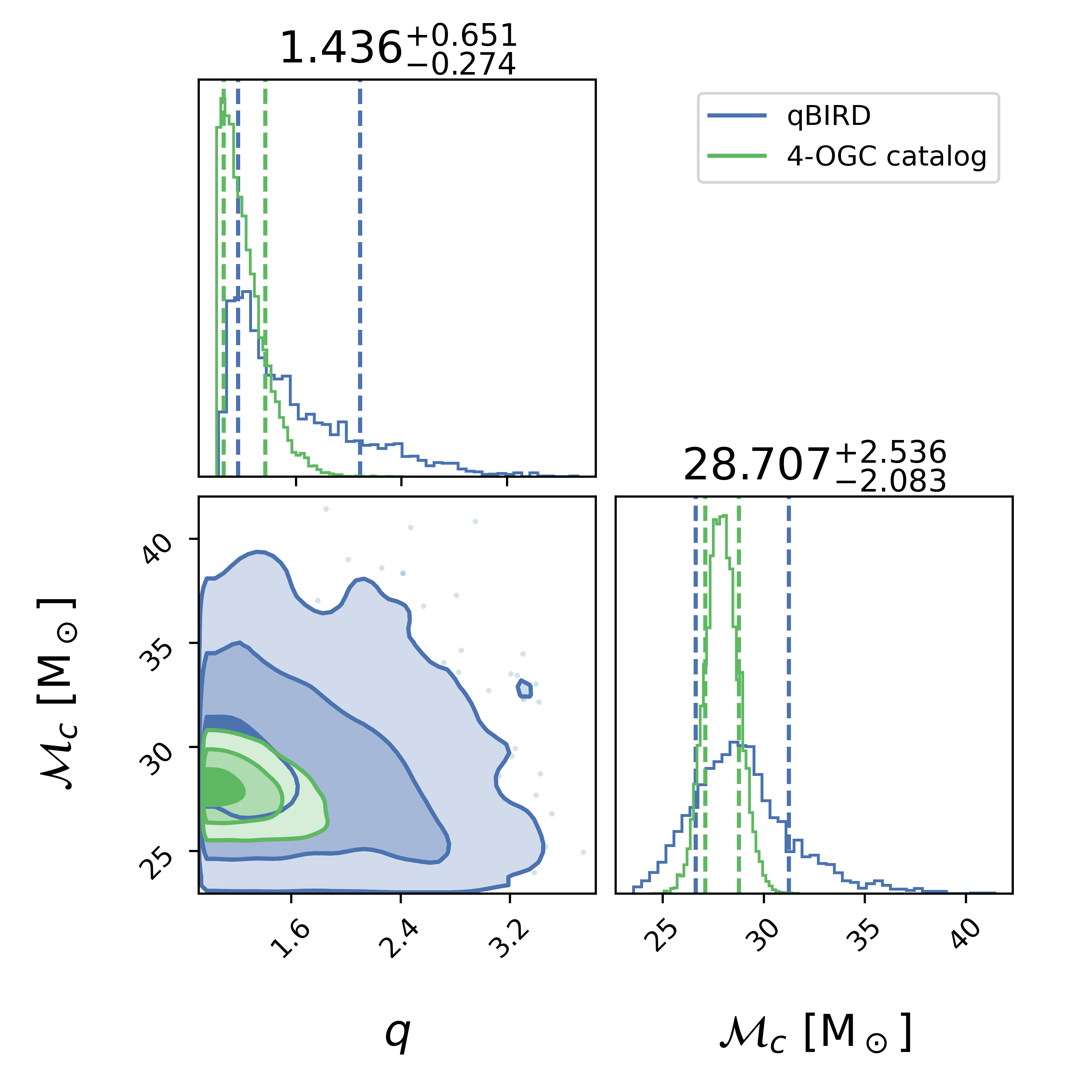}
    \caption{Posterior distributions for chirp mass $\mathcal{M}_c$ and mass ratio $q$ of the GW150914 event inferred using \qbird in blue, compared with results obtained from \pycbc 4-OGC Catalog inference \cite{nitz2023} in green.}
    \label{fig:gw150914_results}
\end{figure*}

We present an additional parameter estimation with \qbird, with 2 parameters estimated, chirp mass $\mathcal{M}_c$ and mass ratio $q$. 
Figure \ref{fig:gw150914_results} shows the posteriors for the first BBH event GW150914, compared to the classical results obtained using 49353 samples from the \pycbc directory in \cite{nitz2023}. In this inference, \qbird used a discretization of $Q = 5$ qubits per parameter and performed 1500 iterations with 4 steps of $W$ in each iteration. In addition, $a = 3$ qubits were used for the ancilla register, leading to a total of 16 qubits to execute the circuit for both cases.

\newpage
Table \ref{tab:tec_details} summarizes the settings for the four parameter estimation scenarios.

\begin{table*}[ht!]
\centering
\caption{A summary of the technical details for the GW inferences in Figures \ref{fig:injections_2param}, \ref{fig:injections_4param}, and \ref{fig:gw150914_results}.}
\label{tab:tec_details}
\begin{tabular}
{l l l l l l l l}
\hline
\hline
\textbf{Inference case} & \textbf{Injected value} & \textbf{Prior} & \textbf{Discr. qubits} & \textbf{$\beta$ schedule} & \textbf{Iteration} & \textbf{W/iter.} \\ \hline
Figure \ref{fig:injections_2param} & \begin{tabular}{l} \hspace{-0.95cm} $\mathcal{M}_c = 19.5\ {\text{M}_{\odot}}$ \\ \hspace{-0.95cm} $q = 2$ \end{tabular} & \begin{tabular}{l}\hspace{-0.95cm} $\mathcal{M}_c \in [19.4, 19.6] \ {\text{M}_{\odot}}$ \\ \hspace{-0.95cm} $q \in [1.9, 2.1]$ \end{tabular} & $Q = 6$& $\beta = 0.5$  & 2100 & 4\\ 
\hline
Figure 
\ref{fig:injections_4param} & \begin{tabular}{l} \hspace{-0.95cm} $\mathcal{M}_c = 30\ {\text{M}_{\odot}}$\\ \hspace{-0.95cm} $q = 0.57$ \\ \hspace{-0.95cm} $d_L = 200 \text{ Mpc}$ \\ \hspace{-0.85cm}$\theta_{jn} = 0.7\ \text{radian}$ \end{tabular} & \begin{tabular}{l} \hspace{-0.95cm} $\mathcal{M}_c \in [27, 35] \ {\text{M}_{\odot}}$ \\ \hspace{-0.95cm} $q \in [0.25, 1]$ \\ \hspace{-0.95cm} $d_L \in [150, 220] \text{ Mpc}$ \\ \hspace{-0.85cm}$\theta_{jn} \in [0, 1]$ radian \end{tabular}& $Q = 3$ & $\beta = 0.05$ & 1200 & 4\\ 
\hline
Figure \ref{fig:gw150914_results} & N/A & \begin{tabular}{l}\hspace{-0.95cm}$\mathcal{M}_c \in [23, 42] \ {\text{M}_{\odot}}$\\ \hspace{-0.95cm} $q \in [1, 4]$ \end{tabular}& $Q = 5$ & $\beta = 0.5$ & 1500 & 4\\ \hline
\hline
\end{tabular}%
\end{table*}

\newpage
\bibliography{bibliography}
\bibliographystyle{aasjournal}

\end{document}